\documentstyle[preprint,eqsecnum,aps,prb]{revtex}

\begin{document}
\title{Thickness dependence of the stability of the charge-ordered state in Pr$%
_{0.5}$Ca$_{0.5}$MnO$_{3}$ thin films.}
\author{W. Prellier, Ch. Simon, A.M. Haghiri-Gosnet, B. Mercey and B. Raveau}
\address{Laboratoire CRISMAT, CNRS\ UMR 6508, ISMRA et Universit\'{e} de Caen, Bd du\\
Mar\'{e}chal Juin, 14050 Caen Cedex, FRANCE.}
\date{\today}
\maketitle

\begin{abstract}
Thin films of the charge-ordered (CO) compound Pr$_{0.5}$Ca$_{0.5}$MnO$_{3}$
have been deposited onto (100)-oriented SrTiO$_{3}$ substrates using the
Pulsed Laser Deposition technique. Magnetization and transport properties
are measured when the thickness of the film is varied. While the thinner
films do not exhibit any temperature induced insulator-metal transition
under an applied magnetic field up to 9T, for thickness larger than 1100\AA\
a 5T magnetic field is sufficient to melt the CO state. For this latest
film, we have measured the temperature-field phase diagram. Compared to the
bulk material, it indicates that the robustness of the CO state in thin
films is strongly depending on the strains and the thickness. We proposed an
explanation based on the distortion of the cell of the film.
\end{abstract}

\newpage

Recently, perovskite type manganites such as RE$_{1-x}$A$_{x}$MnO$_{3}$ (RE
and A are respectively a rare-earth and an alkaline-earth ion), have renewed
interest due to their properties of colossal magnetoresistance (CMR), a huge
decrease in resistance when applying a magnetic field\cite
{VonH,Chahara,Mc,Venky,Jin}. Since most of the technological applications
require thin films, it is essential to understand the effects of the
substrate-induced strains on the properties of these manganites. These
materials are very sensitive to the strains and even very small
perturbations may result in observable effects on the properties. These
effects have indeed been studied experimentally on various compounds \cite
{Amlan2000,Wil,Wang,ODonnel,Sun}. Moreover, among all the properties of
these manganite materials, the phenomenon of charge-ordering (CO) is
probably one of the most remarkable effects. It appears for certain value of
x and particular average A-site cation radi (%
\mbox{$<$}%
$r_{A}$%
\mbox{$>$}%
). It corresponds to an ordering of the charges in two different Mn
sublattices. In fact, the metallic state becomes unstable, below a certain
temperature ($T_{CO}$) and the material goes to an insulating state. $T_{CO}$
decreases with increasing field and the insulating CO state can be totally
suppressed by the application of an external magnetic field \cite
{Sun,Wollan,Kuwahara}. As an example, a 25T magnetic field is required to
melt the CO state in the bulk Pr$_{0.5}$Ca$_{0.5}$MnO$_{3}$ \cite{Tokunaga}.

But, we have shown recently that, when this compound is synthesized as a
thin film on SrTiO$_{3}$ (i.e. with a tensile stress), the melting magnetic
field is reduced to 6T\ \cite{Wil2}, whereas a value of 9T is not sufficient
when the film is grown on LaAlO$_{3}$ to collapse this metastable state \cite
{AMHG}. This means that uniaxial strain (\symbol{126}1-2\% in the plane)
leads to dramatic changes as compared to the bulk material. Following the
same idea, one would expect that the relaxation of the strains along the
direction normal to the plane of the substrate (i.e when the thickness of
the film is varying) should also induce changes in the electronic
properties. For these reasons, it is also interesting to vary the thickness
of the films grown on SrTiO$_{3}$.

In the present work, we have studied the effects of the thickness and the
strains upon the structural and physical properties of Pr$_{0.5}$Ca$_{0.5}$%
MnO$_{3}$ (PCMO) thin films grown on SrTiO$_{3}$ using the Pulsed Laser
Deposition technique. We have particularly considered the changes in lattice
parameters, transport measurements and magnetization. On the basis of these
results, we have determined a temperature-field phase diagram for CO thin
films and compared it to the bulk material.

Thin films of PCMO were grown in-situ using the Pulsed Laser Deposition
(PLD) technique on [100]-SrTiO$_{3}$ substrates (cubic with a=3.905\AA ).
Detailed optimization of the growth procedure was completed and described
previously \cite{Wil2,AMHG}. The structural study was carried out by X-Ray
diffraction (XRD) using a Seifert XRD 3000P for the $\theta -2\theta $ scans
and a Philips MRD X'pert for the in-plane measurements (Cu K$\alpha $, $%
\lambda $=1.5406\AA ). The in-plane parameters were obtained from the (103)$%
_{C}$ reflection (where c refers to the ideal cubic perovskite cell). Direct
current (dc) resistivity ($\rho $) was measured by a four-probe method with
a Quantum Design PPMS and magnetization ($M$) was recorded using a Quantum
Design MPMS SQUID magnetometer as a function of the temperature ($T$) and
the magnetic field ($H$). The composition of the films was checked by
energy-dispersive scattering analyses. It is homogenous and corresponds
exactly to the composition of the target (i.e. Pr$_{0.5}$Ca$_{0.5}$Mn) in
the limit of the accuracy.

The structure of bulk Pr$_{0.5}$Ca$_{0.5}$MnO$_{3}$ is orthorhombic ($Pnma$)
with a=5.395\AA , b=7.612\AA\ and c=5.403\AA\ \cite{Jirak}. Fig. 1 shows a
typical $\theta -2\theta $ scan recorded for a film of (750\AA ). Note that
the same pattern is obtained for all of the films. As already reported, the
film is a single phase, [010]-oriented, i.e. with the [010] axis
perpendicular to the substrate plane\cite{Wil2}. This surprising orientation
results from the lattice mismatch. Indeed, the mismatch (s) between the film
and the substrate can be evaluated using the formula $\sigma =100\ast
(\sigma _{S}-\sigma _{F})/\sigma _{S}$ (where $a_{S}$ and $a_{F}$
respectively refer to the lattice parameter of the substrate and the film).
The smaller mismatch on LaAlO$_{3}$ is obtained for a [010]-axis in the
plane ($\sigma _{LAO}=$-0.4\%), i.e. [101]-axis perpendicular to the
substrate plane. In contrast, the smaller mismatch on SrTiO$_{3}$ ($\sigma
_{STO}=$2.2\%) is found for a [101]-axis in the plane and thus a [010]-axis
normal to the surface of the substrate as found experimentally \cite{Wil2}.
The sharp peaks at 90%
${{}^\circ}$%
intervals in the $\phi $-scan of the PCMO film (see inset of Fig.1) make
evident the existence of a complete in-plane texture of the film. From the
XRD, the out-of-plane parameter of this 750\AA\ film is calculated to be
3.79\AA . However, as a consequence of the biaxial stains, the lattice
parameter is slightly different when the thickness of the film is changing.
Similar results are obtained for the in-plane lattice parameter and the
corresponding data are plotted in Fig.2. When the film is very thin (150\AA
), the in-plane parameter is nearly equal to the one of the substrate
(3.88\AA\ vs 3.9\AA\ for STO). When the thickness of the film is increasing,
the in-plane parameter is decreasing whereas in the mean time, the
out-of-plane parameter is increasing. For large thickness (%
\mbox{$>$}%
1200\AA ), both lattice parameters tend towards a value close to the bulk.
However, they never reach exactly the bulk values and the film is never
fully relaxed. Note that the cell volume is preserved constant and equal to
that of the bulk (\symbol{126}56\AA $^{3}$ referring to the perovskite
cell). Moreover, the temperature dependence of these cell parameters should
be very different from that of bulk samples. It was shown previously that
the orthorhombic distortion induced by the CO ordering is strongly reduced
by the presence of the substrate, preventing the locking of the q value of
the CO on the 
$\frac12$%
value contrary to the bulk sample of the same composition, leading to $q$%
=0.45 for the thicker film \cite{Wil2}.

Fig.3 shows the temperature dependence with different applied magnetic
fields (0, 5T and 9T) for three films thickness. In the absence of applied
field, one always observes a semiconducting behavior with an anomaly around
225K corresponding to the charge-ordered transition (this value is close to
the one found in the bulk material \cite{Tokunaga}). On the contrary, the
results are drastically different under application of external magnetic
fields. Application of a magnetic field up to our maximum value (9T) has
almost no effect on the thinner film (150\AA ): the film is still
semiconducting and the magnetoresistance is very small (Fig.3a). For the
intermediate thickness (750\AA ), while a 5T magnetic field keeps the film
semiconducting, a 9T renders it metallic with a $T_{MI}$ close to 125K (see
Fig.3b). This feature is a typical characteristic of the melting of the CO
state. On the thicker film (1100\AA ), the melting magnetic field is
reducing to 5T (Fig.3c). The thickness dependence of the properties suggests
that the strains play an important role in determining the melting magnetic
field. When the thickness of the film is increasing, there is a reduction of
the strain on the film at room temperature. However, the properties of the
film do not approach those of the bulk as the substrate-induced strain
reduces. In PCMO single crystal, with a 6T magnetic field, the material
remains insulating, whereas the 1100\AA\ film becomes metallic \cite{Tomioka}%
.

The magnetic field dependence of the resistivity at various temperatures is
shown in Fig. 4. The resistivity shows an important decrease on a
logarithmic scale at a critical field ($H_{C}$) indicating the field-induced
melting of the CO state. This field-induced insulator to metal transition,
which accompanies the collapsing of the CO state takes place below $T_{CO}$.
The critical fields in the field-increasing scan and in the field decreasing
scan will respectively be represented by $H_{c}^{+}$ and $H_{c}^{-}$. This
decrease can be viewed as CMR of about five orders of magnitude at 75K. A
clear hysteresis (between the lower and the upper critical fields) is seen
at these temperatures as previously reported on several CO compounds \cite
{Kuwahara,Tomioka}. This hysteretic region is more pronounced when the
temperature is decreasing. The temperature dependence of the large
hysteresis region is a feature of a first order transition and has been
extensively studied for the composition Nd$_{0.5}$Sr$_{0.5}$MnO$_{3}$ in
ref. \cite{Kuwahara}. Similar results can be observed in magnetization
measurements (fig. 5), though our maximum field is then 5T. From the
resistivity measurements, we can deduce a phase diagram for the 1100\AA\
film (Fig. 6). Two remarks should be pointed out. First, as already
mentioned, the transition of the metastable state (i.e. a field-induced
metallic state) is easier or requires a lower field in case of a 1100\AA\
thin film than in the bulk\ \cite{Tomioka}. Second, the shape of the $H-T$
phase diagram is totally different as compared to the bulk single crystal Pr$%
_{0.5}$Ca$_{0.5}$MnO$_{3}$. In fact, this phase diagram of the 1100\AA\ film
has almost the same profile as Pr$_{1-x}$Ca$_{x}$MnO$_{3}$ single crystal
with $0.35<x<0.45$. The hysteretic region grows when the temperature is
decreasing and rapidly vanishes under 8T. This can be explained by the
variation of the q vector. Indeed, the value of the q vector of the CO
modulation (0.45) is also similar to that of Pr$_{1-x}$Ca$_{x}$MnO$_{3}$
single crystal with $0.35<x<0.45$, suggesting that the whole electronic
structure is similar, though the charge transfer is different (x=0.5 instead
of 0.4 in Pr$_{1-x}$Ca$_{x}$MnO$_{3}$). This was discussed in more detail in
a previous paper \cite{AMHG}. In spite of this, the important new result is
the thickness dependence of the film properties. The distortion induced by
the CO cannot fully develop due to the strains imposed by the substrate.
Consequently, the CO is limited to an incommensurate value leading to a less
stable CO state, more sensitive to the magnetic field. In fact, a large CO
gap in the density of states at the Fermi level is associated with a very
stable CO state \cite{Amlan1998}. As the thickness is decreasing, the
in-plane cell parameter is going closer to the substrate value and it can be
supposed that it also follows its temperature dependence leading to a large
cell distortion. Thus, the substrate-induced strain can be measured by the
lattice-distortion $D$ of the film cell defined as the ratio between the
in-plane and the out-of-plane parameters. The evolution of $D$ vs thickness
is plotted in the inset of Fig.2. As the lattice-distortion or equivalently
the tolerance factor \cite{Gold} is approaching one, the distortion of the
perovskite cell is decreasing. In the mean time, the critical field required
to melt the CO state is also decreasing. This behavior is similar to that
observed for bulk materials RE$_{1-x}$Ca$_{x}$MnO$_{3}$, whose CO state is
destabilized when the tolerance factor is getting close to one (or the size
of the A-site cation is increasing) \cite{Rao,Arulja}. In return, the
present evolution of the magnetotransport properties cannot be directly
compared to that of the bulk compound Pr$_{0.5}$Ca$_{0.5}$MnO$_{3}$, due to
the different nature of the cell distortion imposed by the substrate.

In summary, we have shown that the melting magnetic field of the insulating
charge-ordered state, in Pr$_{0.5}$Ca$_{0.5}$MnO$_{3}$ films, is strongly
sensitive to the thickness of the film. While, the application of a magnetic
field up to 9T has almost no effect on the resistivity on a very thin film,
a 5T magnetic field can completely melt the CO state of a 1100\AA\ thick
film. The reduction of this critical magnetic field with the thickness can
be explained by the distortion of the cell of the thin film. Finally, it
appears, that an increase of the thickness in thin films under tensile
strain has the same effect as decreasing the size of the A-site cation in
bulk material: the stability of the CO state is decreasing in Ln$_{0.5}$Ca$%
_{0.5}$MnO$_{3}$ when Ln goes from Y to La \cite{Respaud}.

The authors thank Dr. A. Maignan and Dr. R. Mahendiran for fruitful
discussions.

\section{\protect\bigskip \newpage Figure Captions}

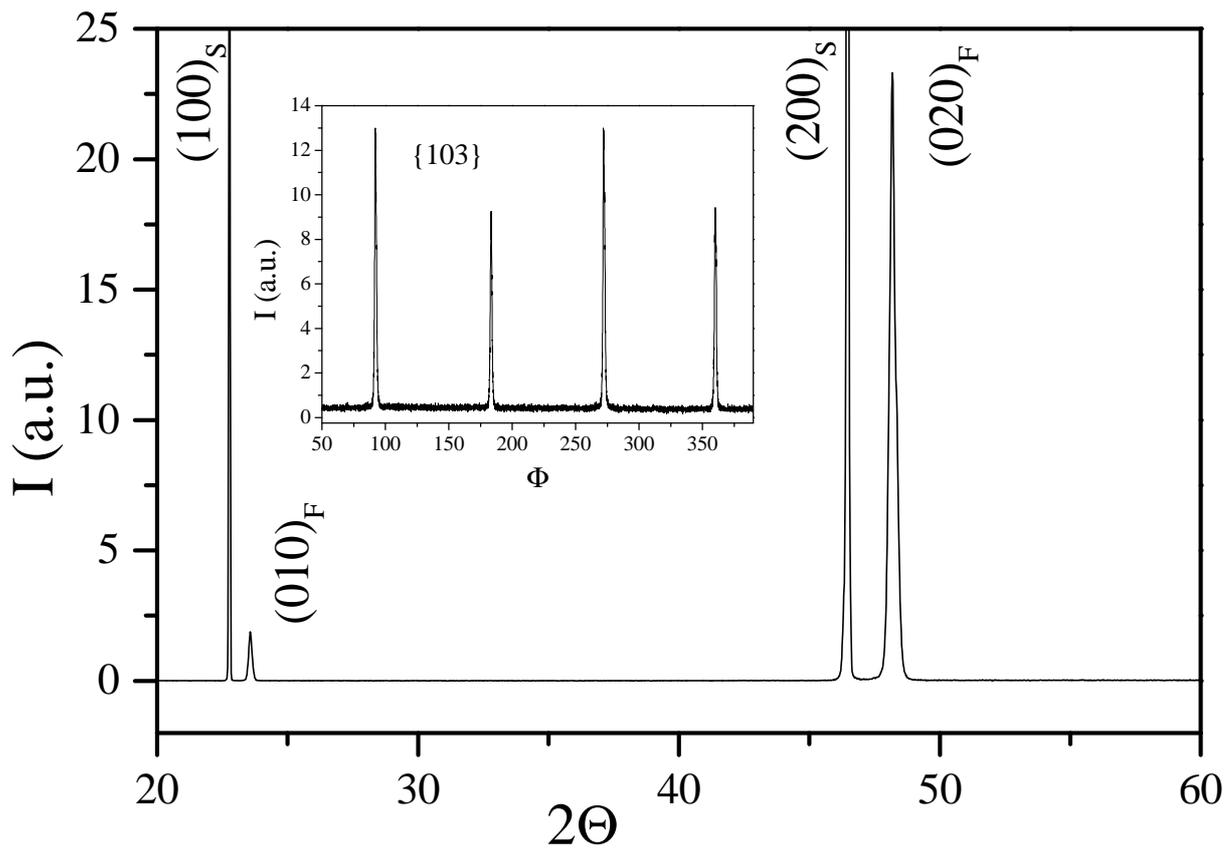
Fig.1 : Room temperature $\theta -2\theta $ XRD pattern of a 750\AA\ film.
Note the high intensity and the sharpness of the peaks. The inset depicts
the $\phi $-scan of the \{103\} family peaks with a four-fold symmetry,
showing the good epitaxy of the film.

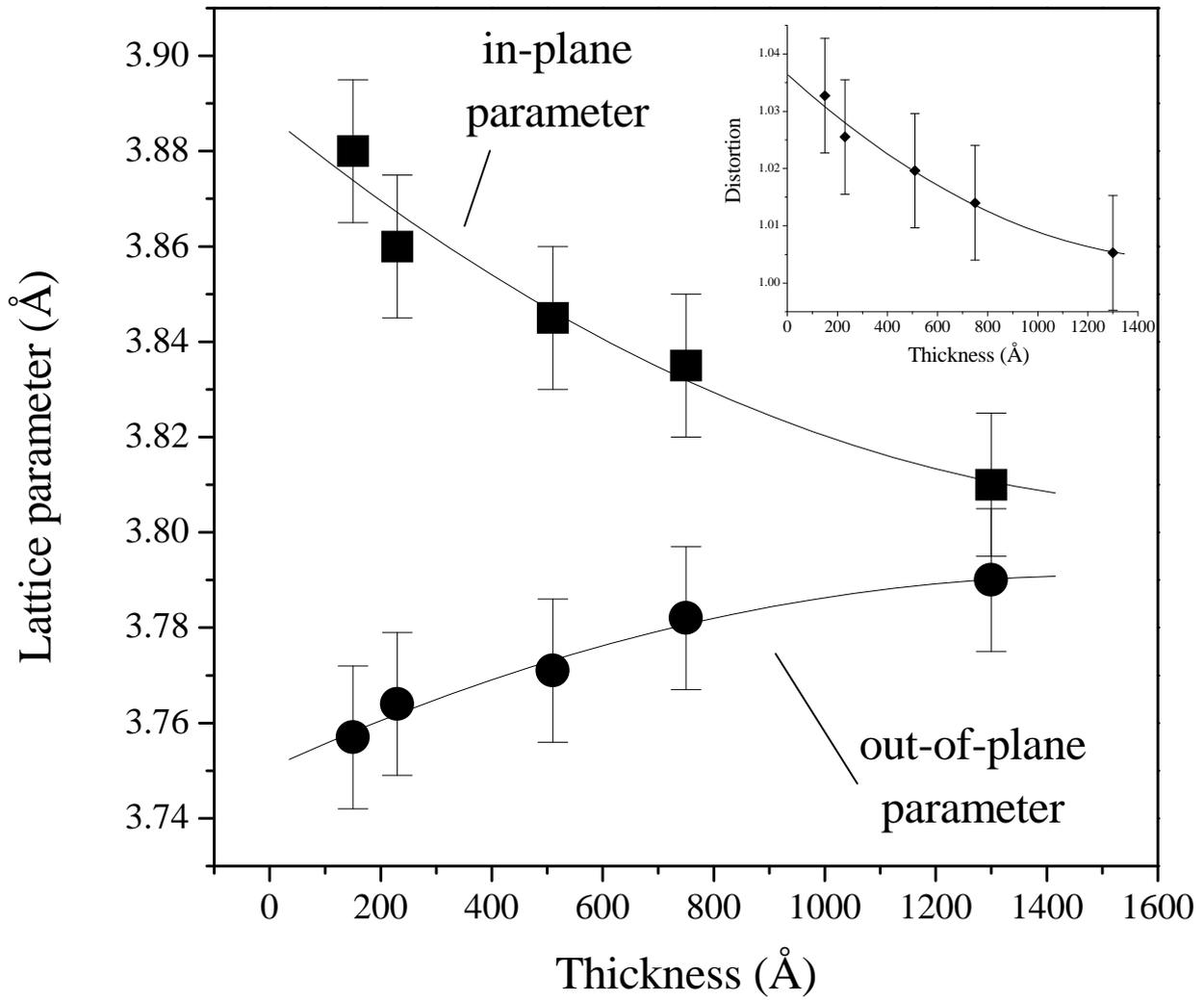
Fig2: Evolution of the in-plane [202] (i.e. 100$_{C}$), and the out-of-plane
parameter [040] (i.e. 001$_{C}$) of PCMO films with different thickness. The
inset depicts the evolution the ratio between the in-plane and the
out-of-plane parameters. Full lines are only a guide for the eyes.

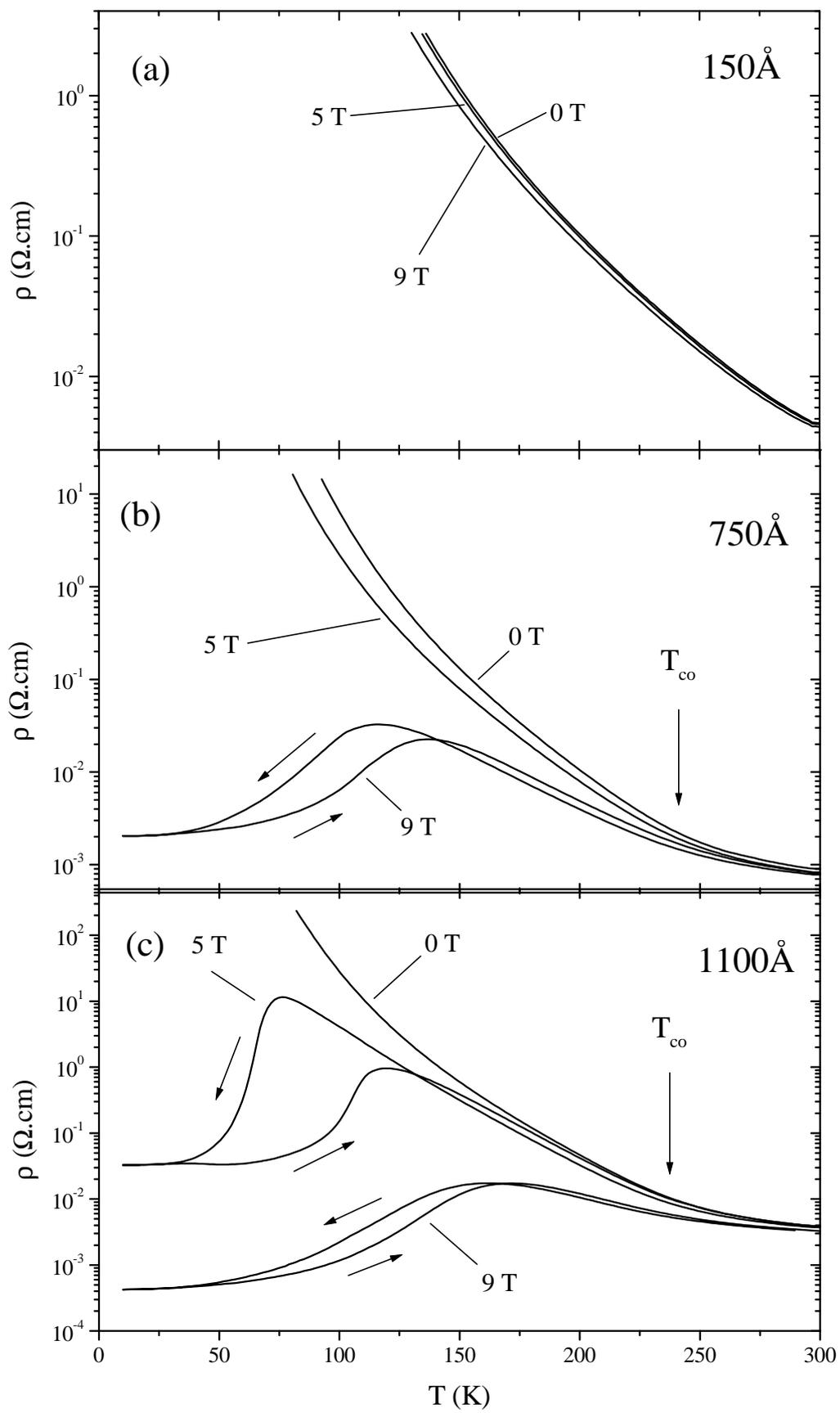
Fig3: $\rho $(T) under different magnetic field (0, 5T and 9T) applied in
the plane of the substrate for different films (a): 150\AA , (b): 750\AA\
and (c): 1100\AA . Arrows indicate the direction of the temperature
variation.

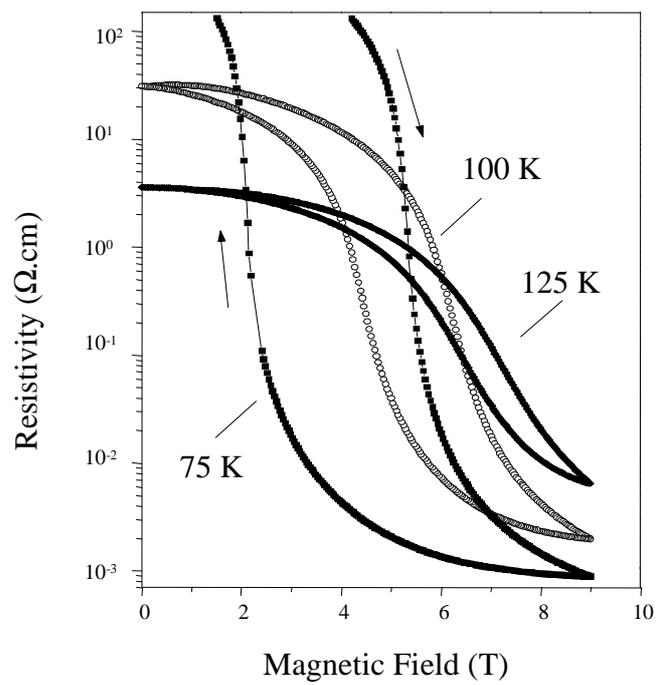
Fig4: Resistivity vs magnetic field at 75K, 100 K and 125K for a 1100\AA\
film. Runs in field-increasing and field-decreasing are denoted by arrows.
Note the large hysteretic region.

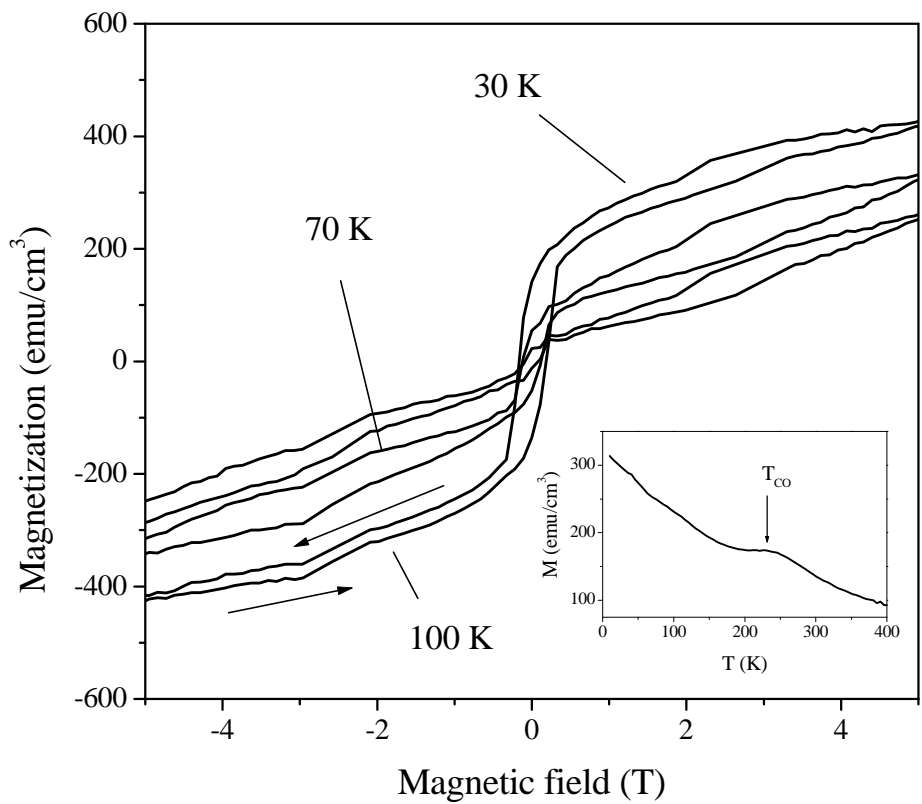
Fig5: In-plane magnetization vs magnetic field at 70K, 100 K and 120K for a
1100\AA\ film. Runs in field-increasing and field-decreasing are denoted by
arrows. The inset shows the magnetization vs temperature recorded under a
magnetic field of 1.45T. The magnetization of the substrate was measured at
each temperature and field and subtracted point by point.

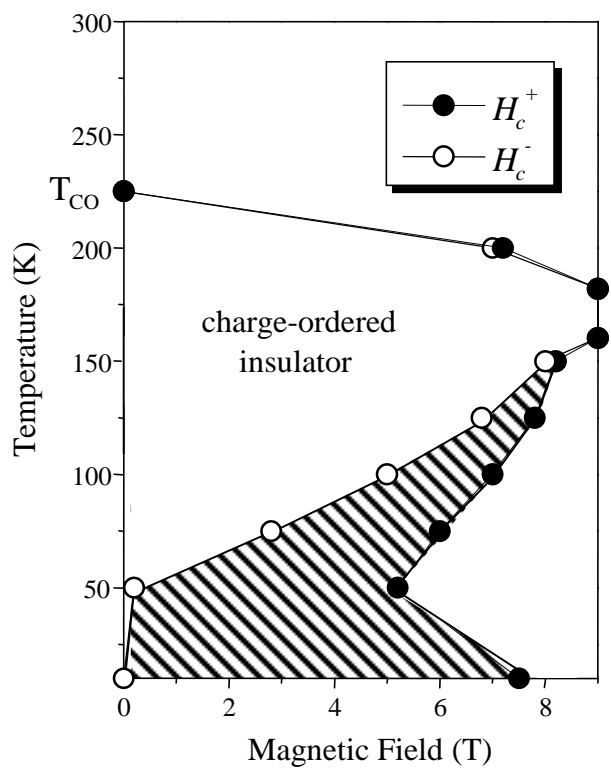
Fig6: Electronic phase diagram of a 1100\AA\ PCMO thin film determined by
the critical fields. $H_{c}^{+}$ and $H_{c}^{-}$ are respectively taken as
the inflexion points in $\rho (H)$ in for up and down sweeps. In the dashed
region, both CO insulating and metallic state are coexisting.

\end{document}